\documentclass[
 aip,
 amsmath,amssymb,
 reprint,%
]{revtex4-1}
\usepackage[makeroom]{cancel}
\usepackage{bm}
\usepackage{amsmath}
\usepackage{mathtools}
\usepackage{amsfonts}
\usepackage{siunitx}
\usepackage{graphicx}

\usepackage{textcomp,booktabs}
\usepackage[usenames,dvipsnames]{xcolor}
\usepackage{colortbl}
\definecolor{FEMGray}{gray}{0.9}
\definecolor{FDGray}{gray}{0.8}

\newcommand{\be}{\begin{equation}}
\newcommand{\ee}{\end{equation}}

\usepackage{hyperref}
\usepackage{adjustbox}

\begin{document}
\begin{minipage}{\textwidth}
The following article has been accepted by Journal of Applied Physics. After it is published, it will be found at this \href{https://aip.scitation.org/journal/jap}{link}.
\end{minipage}
\title{Micromagnetic frequency-domain simulation methods for magnonic systems}
\author{Massimiliano d'Aquino}
\affiliation{Department of Electrical Engineering and ICT, University of Naples Federico II, I-80125 Naples, Italy}
\email{mdaquino@unina.it}
\author{Riccardo Hertel}
\affiliation{Universit{\'e} de Strasbourg, CNRS, Institut de Physique et Chimie des Mat{\'e}riaux de Strasbourg, F-67000 Strasbourg, France}
\email{riccardo.hertel@ipcms.unistra.fr}

\date{Jan. 15th, 2023 (revised version). Original version submitted to J. Appl. Phys. on Oct. 25th, 2022}
\begin{abstract}
We present efficient numerical methods for the simulation of small magnetization oscillations in three-dimensional micromagnetic systems. Magnetization dynamics is described by the Landau-Lifshitz-Gilbert (LLG) equation, linearized in the frequency domain around a generic equilibrium configuration, and formulated in a special operator form that allows leveraging large-scale techniques commonly used to evaluate the effective field in time-domain micromagnetic simulations. By using this formulation, we derive numerical algorithms to compute the free magnetization oscillations (i.e., spin wave eigenmodes) as well as magnetization oscillations driven by ac radio-frequency fields for arbitrarily shaped nanomagnets. Moreover,  semi-analytical perturbation techniques based on the computation of a reduced set of eigenmodes are provided for fast evaluation of magnetization frequency response and absorption spectra as a function of damping and ac field. 
We present both finite difference and finite element implementations and demonstrate their effectiveness on a test case.
These techniques open the possibility to study generic magnonic systems discretized with several hundred thousand (or even millions) of computational cells in a reasonably short time.
\end{abstract}
\maketitle

\section{Introduction}
Research in magnonics \cite{kruglyak_magnonics_2010} aims to exploit the dynamic excitation of a magnetic system to transfer and process information in nanoscale devices. More generally, it addresses the properties of high-frequency magnetic modes and spin waves, alongside possibilities to generate, analyze, manipulate, and exploit such magnetic oscillations. A significant advantage of using spin waves, rather than electrical currents,  for information processing on the nanoscale, is the absence of Ohmic losses and Joule heating. Today, almost twenty years after the first proposal 
to use spin waves for logical operations 
\cite{hertel_domain-wall_2004}, the field has steadily attracted increasing interest and made significant progress in both experiments and theory \cite{chumak_roadmap_2022}.

Theoretical approaches for the determination of small magnetization oscillations trace back to pioneering analytical approaches\cite{Walker1957,Aharoni1991,brown_micromagnetics_1963,mills2007} which were limited to saturated magnetic particles with special shapes. For systems with arbitrary geometry and spatially inhomogeneous magnetization, one has to resort to methods based on numerical simulation. In this respect, problems involving small magnetization oscillations can be addressed by either using time-domain or frequency-domain techniques \cite{baker_proposal_2017}.

The former approach is based on micromagnetic simulations, which reliably allow interpreting and predicting the behavior of the magnetization in ferromagnetic systems on relevant time and length scales, which are usually in the sub-micron and \si{\giga\hertz} range. 
Although they have consistently demonstrated high efficiency and accuracy, micromagnetic simulations of magnonic systems can be time-consuming and tedious. The traditional approach consists of calculating the oscillatory magnetization dynamics generated by an external stimulus over an extended time and, subsequently, analyzing the computed oscillations \cite{mcmichael_magnetic_2005,yan_calculations_2007}. Typical simulations may involve hundreds of thousands of discretization cells, extend over a few tens of nanoseconds, and require time steps in the sub-picosecond range. Relatively complicated Fourier analysis methods are often needed to extract essential information from the simulation results\cite{dvornik_micromagnetic_2013}, which may call for considerable resources in the case of large-scale simulations. 

The above issues can in principle be circumvented by using frequency-domain techniques that are based on the formalism referred to as dynamical matrix method \cite{Born1954,labbe_microwave_1999,vukadinovic_ferromagnetic_2001,grimsditch_magnetic_2004,rivkin_dynamic_2007}.
This method consists of a linearization of the magnetization dynamics around a stable equilibrium configuration  and solution of the resulting eigenvalue problem in order to compute the natural oscillation modes and frequencies of the system.
The main limit of this approach is the intrinsic need to compute and store the dynamical matrix of the system in computer memory. Such matrix, for magnetic systems, is fully-populated owing to long-range magnetostatic interactions, and its dimension scales as $\mathcal{O}(N^2)$, where $N$ is the number of computational cells. 

This structural limit prevents the use of dynamical matrix methods for analyzing magnonic systems, which do require large-scale computations with hundreds of thousand of computational cells.
Furthermore, experiments on magnonic devices are typically performed by measuring the frequency response of the magnetization dynamics driven by suitable radio-frequency (rf) fields produced by microwave antennas. Thus, efficient numerical techniques to compute such forced response are also desirable. 

In this paper, we propose frequency-domain numerical simulation algorithms to study high-frequency magnetic oscillations in arbitrarily shaped ferromagnetic nanostructures. The proposed methods are based on the formulation of linear magnetization dynamics described in ref. \cite{daquino_novel_2009}, which leverages the use of fast large-scale techniques commonly adopted in time-domain micromagnetic solvers. The latter formulation allows implementations based on both finite difference\cite{daquino_computation_2008} (FD) and finite element\cite{forestiere2009finite,baker_proposal_2017} (FEM) spatial discretizations and has been recently used to study magnetization dynamics driven by thermal fluctuations\cite{bruckner2019large} and extended to include nonlinear effects\cite{perna2022computational}. 

The first tool developed in this paper is a large-scale eigensolver which is instrumental to determine the fundamental oscillation modes (termed normal modes or eigenmodes) and frequencies of a ferromagnetic system in the lossless (conservative, zero damping) limit. 
Based on the principles and methods developed for the eigensolver, we then build a linear response solver that allows computing the damped ac steady-state oscillatory magnetization dynamics resulting from a weak externally applied sinusoidal magnetic field.
Finally, by using appropriate perturbation techniques, we develop a semi-analytical theory for straightforward and fast computation of all possible rf-driven magnetization dynamics of a given system as function of (small) damping and arbitrary rf-field. In particular, analytical formulas are provided to obtain the ac steady-state magnetization oscillation vector field, power spectrum and absorbed rf-power from the knowledge of a reduced set of magnetic eigenmodes.
The proposed algorithms exploit general properties of continuum linear magnetization dynamics which do not depend on the choice of spatial discretization. We present both finite difference and finite element implementations of the methods and compare results obtained for an illustrative example.

These methods allow calculating the magnetic modes and their field dependence significantly faster than traditional time-domain micromagnetic simulators. Crucially, we developed our micromagnetic algorithms intending to enable matrix-free large-scale computations of magnetic systems of arbitrary geometry. Accelerated matrix-free techniques are commonly used for the computation of electromagnetic fields via integral formulations \cite{harrington_field_1968,hackbusch_hierarchical_2015}.   
For finite difference solvers, we achieve this by employing a Fast Fourier Transform (FFT) accelerated computation (with $\mathcal{O}(N)$ storage and $\mathcal{O}(N\log N)$ computational cost) of the magnetostatic (demagnetizing) field \cite{yuan_fast_1992}. 
Conversely, for the finite element method, which naturally embeds geometric flexibility and is better suited for curved boundaries,
we employ the well-known hybrid finite/boundary element algorithm by Fredkin and Koehler \cite{fredkin_hybrid_1990} for the magnetostatic field calculation and reduce  its computational costs to a
nearly linear $\mathcal{O}(N)$ scaling through modern $\mathcal{H}2$-matrix compression techniques
\cite{hertel_large-scale_2019}.
In both cases, FD and FEM, our matrix-free implementations of the frequency-domain--based algorithms preserve these advantageous scaling properties by directly using the routines or classes already incorporated in the respective micromagnetic codes without introducing the huge numerical costs that otherwise would arise if the corresponding  ``dynamical matrix'' was set up.
Moreover, our FEM implementation exploits efficient parallelization and optional graphical processing unit (GPU) acceleration, thereby further facilitating large-scale computations. 

\section{Linearized Magnetization Dynamics}
Usual situations in magnonics concern the study of high-frequency small-amplitude modulations $\delta\bm{M}$ of an equilibrium magnetization structure $\bm{M}_0$ in a confined system with nano- or micro-scale dimensions. Magnetization can be decomposed into a static and a small dynamic component 
\be\label{delta_m}
\bm{m}(\bm{x},t)=\bm{m}_0(\bm{x})+\delta\bm{m}(\bm{x}, t) \quad
\text{with} \quad\left|\delta\bm{m}\right|\ll\left|\bm{m}_0\right|\,,
\ee
where $\bm{m}=\bm{M}/M_s$ is the reduced (normalized and dimensionless) magnetization, $\delta\bm{m}=\delta\bm{M}/M_s$, and $M_s$ is the spontaneous magnetization. Due to the micromagnetic nonlinear constraint $\left|\bm{m}\right|=1$, at first-order the deviation $\delta\bm{m}$ must fulfill the condition $\bm{m}_0\cdot\delta\bm{m}=0$.
The static part $\bm{m}_0$, representing an equilibrium
 state, is characterized by a vanishing magnetic torque 
\be\label{notorque}
\bm{m}_0(\bm{x})\times\bm{h}_\text{eff0}(\bm{x})=\bm{0}\quad\forall \bm{x}
\ee 
where 
\be
\bm{h}_\text{eff0}=-\frac{1}{\mu_0M_s}\frac{\delta E[\bm{M}_0(\bm{x})]}{\delta\bm{M}}
\ee
is the reduced micromagnetic effective field \cite{brown_micromagnetics_1963} associated with the magnetic configuration $\bm{M}_0$, defined {\em via} the variational derivative of the micromagnetic energy functional $E[\bm{M}(\bm{x})]$ with respect to $\bm{M}$, and $\mu_0$ is the vacuum permeability.  The effective field contains contributions from all micromagnetic energy terms, in particular from the ferromagnetic exchange $\bm{h}_\text{exc}$, the magneto-crystalline anisotropy $\bm{h}_\text{ani}$, the magnetostatic (demagnetizing) field $\bm{h}_\text{dem}$ and an externally applied field $\bm{h}_\text{ext}$. The vector field $\bm m_0(\bm x)$ fulfills the natural boundary conditions (e.g. $\partial\bm m_0/\partial \bm n=0$ for classical exchange) at the body surface.
Any variation $\delta\bm{m}$ of $\bm{m}_0$ results in a small change in the effective field, such that 
\be \label{delta_h}
\bm{h}_\text{eff}(\bm{x},t)=\bm{h}_\text{eff0}(\bm{x})+\delta\bm{h_\text{eff}}(\bm{x}, t) \quad
\text{with} \quad\left|\delta\bm{h}_\text{eff}\right|\ll\left|\bm{h}_\text{eff0}\right|.
\ee
These assumptions imply that $\delta\bm h_\text{eff}$ is related to the magnetization perturbation vector field  $\delta\bm m$ in a linear fashion:
\begin{equation}\label{eq:C operator}
    \delta\bm h_\text{eff}[\delta\bm m]=-\mathcal{C}\delta\bm m \,.
\end{equation}
When including, for instance, exchange, uniaxial anisotropy and magnetostatics, the operator $\mathcal{C}$ is self-adjoint in the appropriate subspace of square-integrable vector fields with respect to the usual inner product\cite{brown_micromagnetics_1963}. 

It follows from eq.~(\ref{notorque}) that $\bm{h}_\text{eff0}$ is collinear with $\bm{m}_0$, which allows to write 
\be\label{h0}
\bm{h}_\text{eff0}=h_0\,\bm{m}_0
\ee 
where $h_0(\bm x)=\bm{h}_\text{eff0}\cdot\bm{m}_0$ is the projection of the reduced effective field on the reduced magnetization.
Furthermore, we assume a harmonic time dependence of $\delta\bm{m}$ and $\delta\bm{h}$,
\begin{align}\label{harmonic_m}
    \delta\bm{m}(\bm{x}, t)&=\Re\left\{\delta\hat{\bm{m}}(\bm{x})\exp(i\omega t) \right\}\\
    \label{harmonic_h}
    \delta\bm{h}_\text{eff}(\bm{x},t)&=\Re\left\{\delta\hat{\bm h}_\text{eff}(\bm{x})\exp(i\omega t) \right\}
\end{align} 
where $\delta\hat{\bm{m}}(\bm{x})$ and $\delta\hat{\bm{h}}_\text{eff}(\bm{x})$ are complex-valued and position-dependent amplitudes. 

The principle of both methods, the eigensolver algorithm and the forced linear response algorithm, which we will describe in the following sections, consists in inserting equations (\ref{delta_m}), (\ref{delta_h}), (\ref{harmonic_m}) and (\ref{harmonic_h}) into the Landau-Lifshitz-Gilbert (LLG) equation, which, in a normalized form, can be written as
\be \label{gilbert_norm}
\frac{\partial\bm{m}}{\partial\tau}=-\bm{m}\times\bm{h}_\text{eff}+\alpha\bm{m}\times\frac{\partial\bm{m}}{\partial\tau},
\ee
with $\tau=\gamma M_s t$, and subsequently retaining only first-order terms ($\gamma$ is the absolute value of the gyromagnetic ratio and $\alpha$ is the Gilbert damping constant). 
The two algorithms use different further approximations and assumptions and are thus discussed separately.

\section{Magnetic Eigenmodes}\label{sect:eigenmodes}

In the sequel, we briefly recall the formulation proposed by d'Aquino et al.\cite{daquino_novel_2009}.
If we neglect damping, setting $\alpha=0$, and assume that there is no time-dependent external field, 
the first-order approximation of the reduced LLG equation takes the form
\be
    \frac{\partial\delta\bm{m}}{\partial\tau}=-\delta\bm{m}\times\bm{h}_\text{eff0}-\bm{m}_0\times\delta\bm{h}_\text{eff}
\ee
which, after inserting the equations described in the previous sections, leads to
\be 
i\omega\delta\hat{\bm{m}}=-\bm{m}_0\times\left(\delta\hat{\bm{h}}_\text{eff}-h_0\delta\hat{\bm{m}}\right)=\bm m_0\times \mathcal{A}_0\delta\bm{\hat{m}}\,,
\ee 
where $\mathcal{A}_0=\mathcal{C}+h_0(\bm x)\mathcal{I}$ (with $\mathcal{I}$ being the appropriate identity operator) is the (Hessian) operator associated with the second-order variation of the micromagnetic free energy\cite{daquino_novel_2009}.
By using eq.\eqref{eq:C operator} and projecting the latter equation on the plane pointwise perpendicular to the equilibrium $\bm m_0(\bm x)$, it has been shown that nontrivial solutions $\bm \varphi$ of the latter equation satisfy the following generalized eigenvalue problem\cite{daquino_novel_2009}:
\begin{equation}\label{eq:generalized eigenvalue problem}
    \mathcal{A}_{0\perp}  \bm \varphi = \omega \mathcal{B}_0\bm \varphi \,,
\end{equation}
where $\mathcal{A}_{0\perp}=\mathcal{P}_\perp\mathcal{A}_0$ (with $\mathcal{P}_\perp$ being the projection operator $\mathcal{P}_\perp=\mathcal{I}-\bm{m}_0\otimes\bm{m}_0$) is a self-adjoint and positive definite operator and $\mathcal{B}_0=-i \Lambda(\bm m_0(\bm x))$ (the operator notation for the cross product $\Lambda(\bm v)\bm w=\bm v\times \bm w$ has been used) is the invertible Hermitian operator acting on vector fields lying on planes pointwise perpendicular to the equilibrium $\bm m_0(\bm x)$.

Due to its general structure, the eigenvalue problem \eqref{eq:generalized eigenvalue problem} has remarkable properties (see ref.\cite{daquino_novel_2009} for details). The eigenfrequencies $\omega_k$ and eigenfunctions $\bm{\varphi}_k$ represent the magnetization resonant oscillation frequencies and natural modes, respectively.

When discretized on a grid of $N$ nodes $(\bm x_j)_{j=1,\ldots,N}$, the eigenvalue problem \eqref{eq:generalized eigenvalue problem} retains the same structure:
\begin{equation} \label{eq:discrete eig problem}
    A_{0\perp}\cdot\underline{\varphi}=\omega B_0\cdot\underline{\varphi} \,,
\end{equation}
provided that $\underline{\varphi} \,\in \mathbb{C}^{2N}$ are column mesh vectors  containing the collection of $N$ column cell vectors $\varphi_j \,\in \mathbb{C}^2$ (each cell vector has 2 complex components). The discretized operators are\cite{daquino_novel_2009}:
\begin{equation}\label{eq:discretized operators}
    A_{0\perp}=R^T \cdot P_{\bm m_0}\cdot (C + H_0) \cdot R \,
\end{equation}
where $R$ is an appropriate block-diagonal rotation matrix from Cartesian components to local orthogonal coordinates in the planes pointwise perpendicular to $\bm m_{0j}=\bm m_0(\bm x_j)$, $P_{\bm m_0}=\text{diag}(I-\bm m_{01}\otimes \bm m_{01},\ldots I-\bm m_{0N}\otimes \bm m_{0N})$ is the discrete projection operator onto the above planes, and $H_0=\text{diag}(h_{01} I,\ldots,h_{0N} I$).

We remark that the rotation operator $R$ is a $3N\times 2N$ block-diagonal (sparse) matrix which guarantees that the eigenfunctions $\underline{\varphi_k}$ have zero component along the equilibrium magnetization. As a consequence of that, the operator $B_0$ reduces to a $2N\times 2N$ (sparse) block-diagonal matrix having each block proportional to the $2\times 2$ Hermitian matrix $[(0,i),(-i,0)]$.

The eigenvalue problem \eqref{eq:discrete eig problem} can be efficiently solved numerically by using Krylov-subspace methods such as Lanczos/Arnoldi techniques provided, for instance, by the ARPACK library\cite{lehoucq_arpack_1998}. The latter algorithms allow to compute eigenvalues and associated eigenvectors incrementally starting from the one with the smallest/largest magnitude. More interestingly, they only require the computation of matrix-vector products involving the operators $A_{0\perp}$ and $B_0$.  

The Hermitian operator $B_0$ is sparse, invertible, and coincident with its inverse, namely $B_0^H\cdot B_0=I$ (the notation $^H$ means Hermitian conjugate), so this is the easy part. Conversely, products involving the operator $A_{0\perp}$ would require full matrix-vector multiplications implying both storage and computational cost scaling as $O(N^2)$, which becomes unfeasible very soon as $N$ grows.

However, the particular structure of the operator $A_{0\perp}$ in the above formulation (see eq.\eqref{eq:discretized operators} allows the implementation of large-scale computations of matrix-vector products $A_{0\perp}\cdot \underline{v}$ by using the same acceleration techniques used to evaluate the effective field in large-scale micromagnetic simulations. In fact, if ones writes the $j$-th block $\bm w_j$ of the product $\underline{w}=A_{0\perp}\cdot \underline{v}$:
\begin{equation}\label{eq:matrix vector product A0perp}
    \bm w_j=\underbrace{R_j^T}_{2\times 3}\cdot \underbrace{(I-\bm m_{0j}\otimes \bm m_{0j})}_{3\times 3} \cdot \underbrace{(-\delta\bm h^\text{eff}_j[R\cdot \underline{v}] + h_{0j} R_j\cdot\bm v_j)}_{3\times 1} \,,
\end{equation}
it is apparent that, in order to evaluate the product, one can first compute the effective field $\underline{\delta h}^\text{eff}[R\cdot \underline{v}]=-C\cdot R\cdot\underline{v}$ produced by the projected vector field $R\cdot\underline{v}$ and subsequently assemble the rest of the (block-sparse) products. 

We stress that this large-scale computational scheme is a general consequence of the problem formulation\cite{daquino_novel_2009} in the form \eqref{eq:discrete eig problem} and does not depend on the particular choice of discretization method (e.g. finite difference, finite element). 

The finite difference algorithms have been implemented in the code {\em MaGICo}\cite{MaGICo}, which performs micromagnetic simulations both in time and frequency domains using fast FFT magnetostatic solvers and geometrical integration techniques\cite{daquino2005geometrical,daquino_novel_2009} preserving the properties of continuum equations, integrated with the ARPACK\cite{lehoucq_arpack_1998} library.   

For our finite-element implementation in this work,  we use the python interface {\em eigs()} provided by the {\em scipy} library to access ARPACK's functionalities. Concerning the computation of the effective field $\underline{\delta h}^\text{eff}$, we extract the operator $C$ from the micromagnetic finite-element algorithm {\em tetmag}, which features highly efficient methods to calculate the micromagnetic effective fields, e.g., by exploiting $\mathcal{H}2$-type hierarchical matrix compression \cite{hertel_large-scale_2019} and GPU acceleration. To make the routines calculating the effective fields accessible to ARPACK, we prepare python bindings to our implementation of the operator $A_{0\perp}$, whose core components are programmed in C++~\cite{Stroustrup1997}, C ~\cite{ritchie88}, and CUDA~\cite{nickolls_scalable_2008}. 

Solving the system \eqref{eq:discrete eig problem} provides a set of frequencies $\omega_k$ with corresponding oscillation amplitudes $\underline{\varphi}_k$. The user can specify the number $n$ of modes that should be computed, which are typically ordered according to their frequency. Once the eigenvalue problem is solved, the position-dependent oscillation profile $\delta\bm{m}_k(\bm{x}, t)$ of each eigenmode with frequency $\omega_k$ can be obtained as $\delta\bm m_k(\bm{x}, t)=\Re\{\bm\varphi_k(\bm x)\exp(i\omega t)\}$ according to eq.~(\eqref{harmonic_m}). 

The root mean square (RMS) amplitude  of magnetization oscillation in the period of oscillation $T_k=2\pi/\omega_k$ is given in each grid node $\bm x_j$ as:
\begin{equation}
    \langle \delta\bm m_k^2(\bm x_j) \rangle = \frac{1}{T_k} \int_0^{T_k} |\delta\bm m_k(\bm x_j,t)|^2 \,dt = \frac{|\bm\varphi_k(\bm x_j)|^2}{2} \,.
\end{equation}
where the notation $\langle\cdot \rangle$ means average over the ac period.
The above numerically-computed oscillation amplitude can be directly compared with the results of experimental observations as, for instance, those performed by using Brillouin Light Scattering (BLS) measurements\cite{Gubbiotti2010}.

We emphasize again that our implementation of the eigenvalue problem \eqref{eq:discrete eig problem} does not include any dense matrix. 
This clarification appears necessary since the central elements of our mathematical framework correspond to those known from so-called ``dynamical matrix'' methods, in which an operator of the type $\mathcal{A}_{0\perp}$ is typically implemented as a dense matrix---the dynamical matrix giving the method its name. In our case, the absence of dense matrices is essential as it allows us to treat large-scale problems of realistic size. Otherwise, the numerical costs of matrix-vector products $A_{0\perp}\cdot \underline{v}$ in terms of computation time and memory requirements would grow quadratically with the number $N$ of discretization points, $\mathcal{O}(N^2)$, and make large-scale simulations impossible. Instead, we achieve a nearly linear $\mathcal{O}(N)$ scaling by using an operator-based, matrix-free formulation. 

\section{RF-field driven high-frequency dynamics}

The eigensolver described in the previous section can be used to calculate the natural oscillation modes of a ferromagnetic object in static equilibrium. It can help identifying the frequency ranges of interest at which the system may oscillate particularly strongly. Moreover, it indicates which regions within the sample are active at specific frequencies. However, the eigensolver does not provide any information on the nature of the external stimulus required to excite these magnetic eigenmodes, the strength of their oscillation, or the system's behavior in frequency ranges outside those of the resonant modes. 

A situation closer to a realistic experimental setup consists in simulating the response of a magnetic system exposed to an externally applied oscillatory magnetic field. In such a setup, one can obtain an overview of the sample's magnetic high-frequency properties by continuously varying external parameters, such as the strength and direction of an applied static field, the frequency of the oscillatory (rf) field driving the dynamics, or the sample's orientation with respect to that of the rf field. Simulating the frequency-dependent absorption spectrum and its changes induced by such parameter variations results in data that, in many cases, can be directly compared with experiments.

In such situations, where the magnetization performs oscillations driven by an externally applied sinusoidal field, the prerequisites that allowed us to linearize the LLG equation remain fulfilled if the amplitude of the applied oscillatory field is sufficiently small. Specifically, we assume a time-harmonic external field that can possibly be spatially inhomogeneous (as that produced by a microstrip antenna): 
\be 
\delta\bm{h}_\text{ext}(\bm x,t)=\Re\left\{\delta\hat{\bm{h}}_\text{ac}(\bm x)\exp(i\omega t)\right\}
\ee 
 with sufficiently small $\left|\delta\bm{h}_\text{ext}\right|$, where the field strength is
expressed in reduced units, $\delta\bm{h}_\text{ext}=\delta\bm{H}_\text{ext}/M_s$. In the perturbative approach to linearizing the LLG equation, the oscillating external field $\delta\bm{h}_\text{ext}$  is considered to be of the same order as the other oscillating components $\delta\bm{m}$ and $\delta\bm{h}_\text{eff}$. Contrary to the assumptions we made for the eigensolver, we consider here a non-vanishing damping constant $\alpha\neq 0$.

Under these assumptions and using the same operator notations as in the previous section, the linear magnetization dynamics can be written in the frequency domain as \cite{daquino_novel_2009}$^,$\footnote{We have recently become aware that a preprint, 
%by Lin and Lomakin, 
posted on the arXiv server after the submission of this article, describes a similar approach \cite{lin_linearized_2022}.
%based on the formulation of ref.\cite{daquino_novel_2009}.
}:
\begin{equation} \label{eq:forced linear LLG dynamics}
    -i\omega\bm m_0\times\delta\hat{\bm{m}}=-\mathcal{P}_\perp\left(\delta\hat{\bm{h}}_\text{eff}-h_0\delta\hat{\bm{m}}\right)-\mathcal{P}_\perp\delta\hat{\bm h}_\text{ac} + i\omega \alpha \delta\hat{\bm{m}}  \,,
%    = \\
 %   \bm m_0\times \mathcal{A}_0\bm \delta\bm \hat{m}\,,
\end{equation}
which can be recast in the following operator form:
\begin{equation}
    \mathcal{L}_0\delta\hat{\bm m} = \mathcal{P}_\perp\delta\hat{\bm h}_\text{ac} \,,
\end{equation}
where $\mathcal{L}_0[\bm m_0,\omega,\alpha]=\mathcal{A}_{0\perp}-\omega B_0+i\alpha\omega\mathcal{I}$. The formal inversion of the operator $\mathcal{L}_0$ provides the magnetization linear response 
\begin{equation}\label{eq:magnetic susceptibility}
    \delta\hat{\bm m}(\bm x)={\chi}_0(\omega,\alpha)\delta\hat{\bm h}_\text{ac}(\bm x)
\end{equation}
in terms of the ac susceptibility operator ${\chi}_0=\mathcal{L}_0^{-1}\mathcal{P}_\perp$ acting on the vector field $\delta\hat{\bm h}_\text{ac}(\bm x)$.

By introducing the same discretization scheme as in the previous section, one arrives to the following linear system:
\begin{equation}\label{eq:discrete forced LLG response}
    L_0\cdot \underline{\delta \hat{m}} = R^T\cdot P_{\bm m_0}\cdot\underline{\delta\hat{h}}_\text{ac} \,,
\end{equation}
with $L_0=(A_{0\perp}-\omega B_0+i\alpha\omega I)$ being a matrix whose inversion provides the magnetization small oscillation field $\underline{\delta\hat{m}}=L_0^{-1}\cdot R^T\cdot P_{\bm m_0}\cdot\underline{\delta\hat{h}}_\text{ac}$ around the equilibrium $\bm m_0(\bm x)$ driven by the time-harmonic external field $\delta\bm h_\text{ext}$. 

The matrix $L_0^{-1}\cdot R^T\cdot P_{\bm m_0}$ is the discrete counterpart of the ac susceptibility operator ${\chi}_0(\omega,\alpha)$ of the magnetization around the equilibrium $\bm m_0$.

Based on the discussion on large-scale implementation of the eigenmodes calculation performed in the previous section, we emphasize that again large-scale inversion of the operator $L_0=(A_{0\perp}-\omega B_0+i\omega\alpha I)$ can be achieved by exploiting Krylov-subspace techniques such as, for instance, Generalized Minimum Residual (GMRES) method\cite{baker_technique_2005} which require only matrix-vector products, and accelerating matrix-vector products $A_{0\perp}\cdot\underline{v}$ using large-scale computation of the micromagnetic effective field according to the decomposition shown in eq.\eqref{eq:matrix vector product A0perp}.

It is worth remarking that, similarly to what happens for the eigenvalue problem \eqref{eq:discrete eig problem}, the large-scale computational scheme expressed by eq.\eqref{eq:discrete forced LLG response} does not depend on the choice of the spatial discretization method (finite differences, finite elements).

We have implemented the method mentioned above, which hereafter we refer to as Matrix-Free Micromagnetic Linear Response Solver (MF-$\mu$LRS), both with FEM and FD discretizations. 
In our finite-element implementation, we use the LGMRES algorithm \cite{baker_technique_2005} provided by python's scipy library \cite{2020SciPy-NMeth}. The operator ${\cal L}_0[\bm m_0,\omega, \alpha]$ is implemented using optimized routines written in C++ and CUDA, taken from our proprietary tetmag code and made accessible to the scipy solver through appropriate python bindings. The FD MF-$\mu$LRS has been implemented within {\em MaGICo}\cite{MaGICo} by inverting eq.\eqref{eq:discrete forced LLG response} via GMRES method with reverse communication directly embedded in the code.
 
 A typical calculation based on the linear response \eqref{eq:discrete forced LLG response} is the power spectrum of magnetization:
 \begin{equation}\label{eq:power spectrum}
     p(\omega)=\frac{1}{V}\int \frac{|\delta\hat{\bm m}(\bm x)|^2}{2}\, dV  \,,
 \end{equation}
where the integral is performed over the magnetic system volume $V$, which reveals the resonant frequencies that match a given excitation field profile $\delta\hat{\bm h}_\text{ac}(\bm x)$ as a function of frequency. This can be numerically evaluated (in dimensionless form) as:
 \begin{equation}\label{eq:discrete power spectrum}
     p(\omega)
     \approx\frac{1}{V}\sum_{j=1}^N \frac{|\delta\hat{\bm m}_j|^2}{2} V_j = \frac{1}{2V}\underline{\delta\hat{m}}^H\cdot V_\text{cell}\cdot\underline{\delta\hat{m}}\,,
    %  \approx\frac{1}{V}\sum_j^N \frac{|\delta\hat{\bm m}_j|^2}{2} V_j = \frac{1}{2N}\underline{\delta\hat{m}}^H\cdot \underline{\delta\hat{m}}\,,
 \end{equation}
where the notation $^H$ means Hermitian conjugate, the sum is extended over all $N$ grid nodes and $V_j, V$ are the volumes occupied by the $j$-th cell, $V=V_1+\ldots+V_N$ is the volume of the whole magnetic system, respectively, and $V_\text{cell}=\text{diag}(V_1,V_1,\ldots,V_N,V_N)$ is a diagonal matrix. The matrix $V_\text{cell}$ is instrumental for the treatment of unstructured grids as it is the case of finite element methods, whereas it reduces to $V_\text{cell}=\frac{V}{N}\,I$ (with $I$ being the $2N\times 2N$ identity matrix) for finite difference discretization.

Another quantity of interest concerning applications is the average power absorbed by the magnetic system under the action of the external rf-field, which in the ac steady state at frequency $\omega$ can be expressed (in dimensionless form) as:
\begin{equation}
    P_\text{abs}(\omega)=\langle \frac{1}{V}\int \delta\bm h_\text{ext}\cdot\frac{\partial\delta{\bm m}}{\partial t} \,dV \rangle
\end{equation}
where the notation $\langle\cdot \rangle$ means average over the ac period. The average power can be computed as $P_\text{abs}(\omega)=\Re\{\hat{P}_\text{abs}(\omega)\}$, with $\hat{P}_\text{abs}$ being the
(complex) magnetic absorbed power:
\begin{equation}\label{eq:absorbed power}
    \hat{P}_\text{abs}(\omega) =\frac{1}{2\,V}\int i\omega \delta\hat{\bm h}_\text{ac}^* \underbrace{\cdot{\chi}_0(\omega,\alpha)\delta\hat{\bm h}_\text{ac}}_{\delta\hat{\bm m}}\,dV \,,
\end{equation}
where eq.\eqref{eq:magnetic susceptibility} has been used.
This can be numerically computed using eq.\eqref{eq:discrete forced LLG response}, for a given frequency $\omega$, as:
\begin{equation}\label{eq:discrete absorbed power}
        % \hat{P}_\text{abs}(\omega)\approx \frac{1}{2N} \,\underline{\delta\hat{h}}_\text{ac}^H\cdot P_{\bm m_0}\cdot R\cdot  i\omega\underline{\delta\hat{m}} = \frac{1}{2N} \underline{\delta\hat{h}}_\text{ac}^H\cdot i\omega X_0 \cdot\underline{\delta\hat{h}}_\text{ac}
                \hat{P}_\text{abs}(\omega)\approx \frac{1}{2V} \,\underline{\delta\hat{h}}_\text{ac}^H\cdot P_{\bm m_0}\cdot R\cdot V_\text{cell} \cdot i\omega\underline{\delta\hat{m}} =  \underline{\delta\hat{h}}_\text{ac}^H\cdot i\omega X_0 \cdot\underline{\delta\hat{h}}_\text{ac}
\end{equation}
with $X_0(\omega,\alpha)=\frac{1}{2V} P_{\bm m_0} \cdot R \cdot V_\text{cell} \cdot L_0^{-1}\cdot R^T\cdot P_{\bm m_0}$ being the discrete counterpart of the volume-weighted susceptibility operator. %$\chi_0(\omega,\alpha)$.

\section{Semi-analytical computation of rf-field driven dynamics for low damping}\label{sect:semi-analytic}

The magnetization's linear response expressed by eq.~\eqref{eq:forced linear LLG dynamics} holds for any value of the damping constant $\alpha$ and rf-field, and accordingly for the discrete relationship~\eqref{eq:discrete forced LLG response}. Thus, in principle, in order to explore a range of damping values and/or rf-fields, the inversion of eq.~\eqref{eq:discrete forced LLG response} must be performed for each different condition, resulting in a computationally intensive task, although feasible due to the large-scale formulation.

Nevertheless, for sufficiently low values of the damping $\alpha$, it is possible to derive an approximate semi-analytical expression of the response $\delta\hat{\bm m}(\bm x)$ by only using a set of normal modes (eigenmodes) computed according to the procedure outlined in section III.

As preliminary step, we recall that, if $(\bm \varphi_k,\omega_k)$ is an eigenpair, then so it is $(\bm \varphi_k^*,-\omega_k)$. Moreover, the eigenmodes $\bm\varphi_k$ satisfy a special orthogonality property\cite{daquino_novel_2009}:
\begin{equation}\label{eq:orthogonality eigenmodes}
    \frac{1}{V}\int \bm\varphi_k^*\cdot\mathcal{A}_{0\perp}\bm\varphi_h \,dV =\delta_{kh}\,
\end{equation}
where the integral is performed over the magnetic system of volume $V$ and the symbol $\delta_{kh}$ denotes Kronecker's delta. We remark that eq.\eqref{eq:orthogonality eigenmodes} differs from the usual $\mathbb{L}^2$ orthogonality. 

The latter property reflects in the fact that one can compute the whole set of orthonormal discrete eigenmodes $\underline{\varphi}_k$ for which it happens:
\begin{equation}\label{eq:orthonormality}
    \Phi^H\cdot A_{0\perp}\cdot\Phi=I \,,
\end{equation}
where the notation $^H$ means Hermitian conjugate and $\Phi$ is the $2N\times 2N$ complex matrix having eigenvectors $\underline{\varphi}_k$ as columns. 
In addition, the generalized eigenvalue problem \eqref{eq:discrete eig problem} implies that:
\begin{equation}\label{eq:spectral theorem}
    A_{0\perp}=B_0\cdot\Phi\cdot\Omega\cdot\Phi^{-1} \,,
\end{equation}
with $\Omega=\text{diag}(-\omega_N,\ldots,\omega_{N})$ being the $2N\times 2N$ diagonal matrix containing all eigenfrequencies.

Now, when small damping $\alpha\ll 1$ is considered, the perturbation technique developed in ref.\cite{daquino_novel_2009} provides the expressions of the perturbed eigenfrequencies $\omega_k'$:
\begin{equation}\label{eq:perturbed eigenfrequencies}
    \omega_k'=\omega_k+\delta\omega_k\,,\,\delta\omega_k=i\,\alpha\omega_k^2||\underline{\varphi}_k||^2_2
\end{equation}
which are accurate to the order $\mathcal{O}(\alpha^2)$ (the notation $||\underline\varphi_k||^2_2=\underline\varphi_k^H\cdot \underline\varphi_k$ has been used to denote the usual 2-norm in $\mathbb{C}^{2N}$). We remark that $\Re\{\delta\omega_k\}=\alpha\omega_k^2||\underline{\varphi}_k||^2_2=1/\tau_k$ have the physical meaning of the decay constants associated with eigenmodes, which tell that the $k$-th eigenmode  practically vanishes after $\sim5\,\tau_k (\gamma M_s)^{-1}$ seconds. 

Let us now rewrite eq.~\eqref{eq:discrete forced LLG response} in the following way:
\begin{equation}\label{eq:discrete forced LLG response 2}
      \underline{\delta \hat{m}} = [B_0\cdot A_{0\perp}-\omega I+i\,\alpha\omega B_0]^{-1}\cdot B_0\cdot R^T\cdot P_{\bm m_0}\cdot\underline{\delta\hat{h}}_\text{ac} \,, 
\end{equation}
where we have used the fact that $B_0^{-1}=B_0$. Now, by using the generalized spectral decomposition~\eqref{eq:spectral theorem}, one has:
\begin{equation}
      \underline{\delta \hat{m}} = [\Phi\cdot\Omega\cdot\Phi^{-1}-\omega I+i\,\alpha\omega B_0]^{-1}\cdot B_0\cdot R^T\cdot P_{\bm m_0}\cdot\underline{\delta\hat{h}}_\text{ac} \,. 
\end{equation}
To the first-order with respect to $\alpha$, one can write:
\begin{equation}
    [\Phi\cdot\Omega\cdot\Phi^{-1}-\omega I+i\,\alpha\omega B_0]^{-1}\approx[\Phi\cdot\Omega'\cdot\Phi^{-1}-\omega I]^{-1} \,,
\end{equation}
with $\Omega'$ being the diagonal matrix with entries $\omega_k'$ given by eq.~\eqref{eq:perturbed eigenfrequencies}. By using the latter equation, eq.~\eqref{eq:discrete forced LLG response 2} can be rewritten as:
\begin{equation}\label{eq:approx discrete forced LLG response}
      \underline{\delta \hat{m}} \approx [\Phi\cdot(\Omega'-\omega I)^{-1}\cdot\Phi^{-1}]\cdot B_0\cdot R^T\cdot P_{\bm m_0}\cdot\underline{\delta\hat{h}}_\text{ac} \,.
\end{equation}
By remembering eqs.~\eqref{eq:orthonormality}-\eqref{eq:spectral theorem}, one obtains: 
\begin{equation}\label{eq:approx discrete forced LLG response2}
      \underline{\delta \hat{m}} \approx [\Phi\cdot(\Omega'-\omega I)^{-1}\cdot\Omega \cdot\Phi^H\cdot B_0^H]\cdot B_0\cdot R^T\cdot P_{\bm m_0}\cdot\underline{\delta\hat{h}}_\text{ac} \,.
\end{equation}
Equation \eqref{eq:approx discrete forced LLG response2} can be regarded as the sum of projections of $\delta\underline{\hat{m}}$ over all the eigenmodes:
\begin{equation}
      \underline{\delta \hat{m}} \approx \left[\sum_k \frac{\omega_k}{\omega_k'-\omega} \underline{\varphi}_k\cdot\underline{\varphi}_k^H \right]\cdot R^T\cdot P_{\bm m_0}\cdot\underline{\delta\hat{h}}_\text{ac} \,,
\end{equation}
or, equivalently, as 
\begin{equation}\label{eq:approx discrete forced LLG response series}
     \underline{\delta \hat{m}} \approx \Phi\cdot\underline{a} ,\quad{a_k}=\frac{\omega_k}{\omega_k'-\omega} \,\underline{\varphi}_k^H \cdot R^T\cdot P_{\bm m_0}\cdot\underline{\delta\hat{h}}_\text{ac} \,,
\end{equation}
with $\underline{a}$ being the (column) vector of the expansion coefficients $a_k$ of $\delta\underline{\hat{m}}$ over the eigenmodes $\underline{\varphi}_k$.

When a reduced set of $n\ll N$ eigenmodes associated with positive eigenfrequencies $0<\omega_1<\ldots<\omega_n$ is computed by using methods described in section \ref{sect:eigenmodes}, one can calculate the approximate linear magnetization response by simply truncating eq.\eqref{eq:approx discrete forced LLG response series}:
\begin{equation}\label{eq:approx reduced discrete forced LLG response}
      \underline{\delta \hat{m}} \approx %\left[
      \sum_{k=1}^n \frac{\omega_k}{\omega_k'-\omega} \,\underline{\varphi}_k %\right]
      \cdot\underline{\varphi}_k^H \cdot R^T\cdot P_{\bm m_0}\cdot\underline{\delta\hat{h}}_\text{ac} \,,
\end{equation}
which is of course a good approximation of \eqref{eq:approx discrete forced LLG response} in the frequency range $[\omega_1,\omega_n]$.

We stress that eq.~\eqref{eq:approx reduced discrete forced LLG response} represents all the possible frequency responses as a function of damping (the dependence on $\alpha$ occurs through  $\omega_k'$ according to eq.~\eqref{eq:perturbed eigenfrequencies}) and rf-field distribution. Thus, provided that the damping is sufficiently low, as is the case of usual magnetic materials, one can easily have access to the frequency-dependent magnetization dynamics driven by %all possible 
any low-amplitude
rf-fields within the entire frequency range spanned by the set of computed normal modes $\underline{\varphi}_1,\ldots,\underline{\varphi}_n$. 

By using eq.\eqref{eq:approx reduced discrete forced LLG response}, the power spectrum \eqref{eq:discrete power spectrum} of magnetization is easily evaluated, as well as the magnetic absorbed power $\eqref{eq:discrete absorbed power}$. 
Furthermore, by neglecting terms due to coupling between different modes $\underline{\varphi_k}$ and $\underline{\varphi_h}$ with $h\neq k$ in \eqref{eq:discrete power spectrum} when using eq.\eqref{eq:approx reduced discrete forced LLG response}, 
one can express the approximate power spectrum $\tilde{p}(\omega)$ as:
\begin{equation}\label{eq:approx power spectrum}
    % \tilde{p}(\omega)=\frac{1}{2N}\sum_{k=1}^n \frac{\omega_k^2\,|h_k|^2\,||\underline{\varphi}_k||_2^2}{(\omega-\omega_k)^2+(\alpha\omega_k^2 \,||\underline{\varphi}_k||_2^2)^2} \,,
    \tilde{p}(\omega)=\frac{1}{2V}\sum_{k=1}^n \frac{\omega_k^2\,|h_k|^2\,||\underline{\varphi}_k||_{V_\text{cell}}^2}{(\omega-\omega_k)^2+(\alpha\omega_k^2 \, ||\underline{\varphi}_k||_2^2)^2 } \,,
\end{equation}
where $h_k=\underline{\varphi_k}^H\cdot R^T\cdot P_{\bm m_0}\cdot\underline{\delta\hat{h}}_\text{ac}$ is the projection of the external rf-field on the $k$-th eigenmode and $||\underline{\varphi}_k||_{V_\text{cell}}^2 = \underline{\varphi}_k^H\cdot V_\text{cell}\cdot \underline{\varphi}_k$.

Finally, by using eq.~\eqref{eq:approx reduced discrete forced LLG response} in eq.~\eqref{eq:discrete absorbed power}, a compact expression for the magnetic absorbed power $\hat{P}_\text{abs}(\omega)$ is readily obtained:
\begin{equation}\label{eq:compact discrete absorbed power}
% \hat{P}_\text{abs}(\omega)=\frac{1}{2N}\sum_{k=1}^n \frac{i\omega\,|h_k|^2 \omega_k}{(\omega_k-\omega)+i\alpha\omega_k^2||\bm\varphi_k||_2^2} \,.
\hat{P}_\text{abs}(\omega)=\frac{1}{2V}\sum_{k=1}^n \frac{i\omega\,h_k'^*\,h_k \omega_k}{(\omega_k-\omega)+i\alpha\omega_k^2||\underline{\varphi_k}||_2^2} \,,
\end{equation}
where $h_k'=\underline{\varphi_k}^H\cdot V_\text{cell} \cdot R^T\cdot P_{\bm m_0}\cdot\underline{\delta\hat{h}}_\text{ac}$ is the volume-weighted projection of the rf-field on the $k$-th eigenmode. 
From the latter equation, the active (i.e., average) power absorbed by the magnetic system under the action of the ac external field can be computed:
\begin{equation}\label{eq:compact discrete absorbed active power}
    P_\text{abs}(\omega)=\Re\{ \hat{P}_\text{abs}(\omega)\} \,,
\end{equation}
which can be very useful for comparisons with ferromagnetic resonance (FMR) measurements in experiments.

If one approximates the cell volume matrix using the average cell volume, i.e. $V_\text{cell}\approx \frac{V}{N}\,I$, it happens that $h'_k=(V/N)h_k$ and a simple expression holds for $P_\text{abs}(\omega)$:
\begin{equation}\label{eq:approx discrete absorbed active power}
P_\text{abs}(\omega)\approx\frac{1}{2N}\sum_{k=1}^n \frac{\alpha\omega |h_k|^2 \omega_k^3||\underline{\varphi}_k||_2^2}{(\omega_k-\omega)^2+(\alpha\omega_k^2||\underline{\varphi}_k||_2^2)^2} \,.
\end{equation}
By observing eqs.~\eqref{eq:approx power spectrum}-\eqref{eq:approx discrete absorbed active power}, we recognize that twice the magnitude $|\delta\omega_k|$ in eq.\eqref{eq:perturbed eigenfrequencies} expresses the full width half maximum linewidth $\Delta f_{k,\text{FWHM}}=2\gamma M_s/(2\pi)\,\alpha\omega_k^2||\underline{\varphi}_k||^2_2$ (in physical units) of the  spectral peak associated with the $k-$th mode and, consequently, the quality factor $Q_k=\omega_k/|2\delta\omega_k|=1/(2\alpha\omega_k||\underline{\varphi}_k||^2_2)$.

It is worthwhile remarking that the above derivations only make use of structural properties of the generalized eigenvalue formulation and, therefore, do not depend on the particular choice of the spatial discretization method.

Thus, equations \eqref{eq:approx reduced discrete forced LLG response},\eqref{eq:approx power spectrum},\eqref{eq:compact discrete absorbed active power} and \eqref{eq:approx discrete absorbed active power} allow for large-scale computation of damped linear magnetization dynamics driven by arbitrary rf-fields.

\begin{figure*}[ht]
\includegraphics[width=\linewidth]{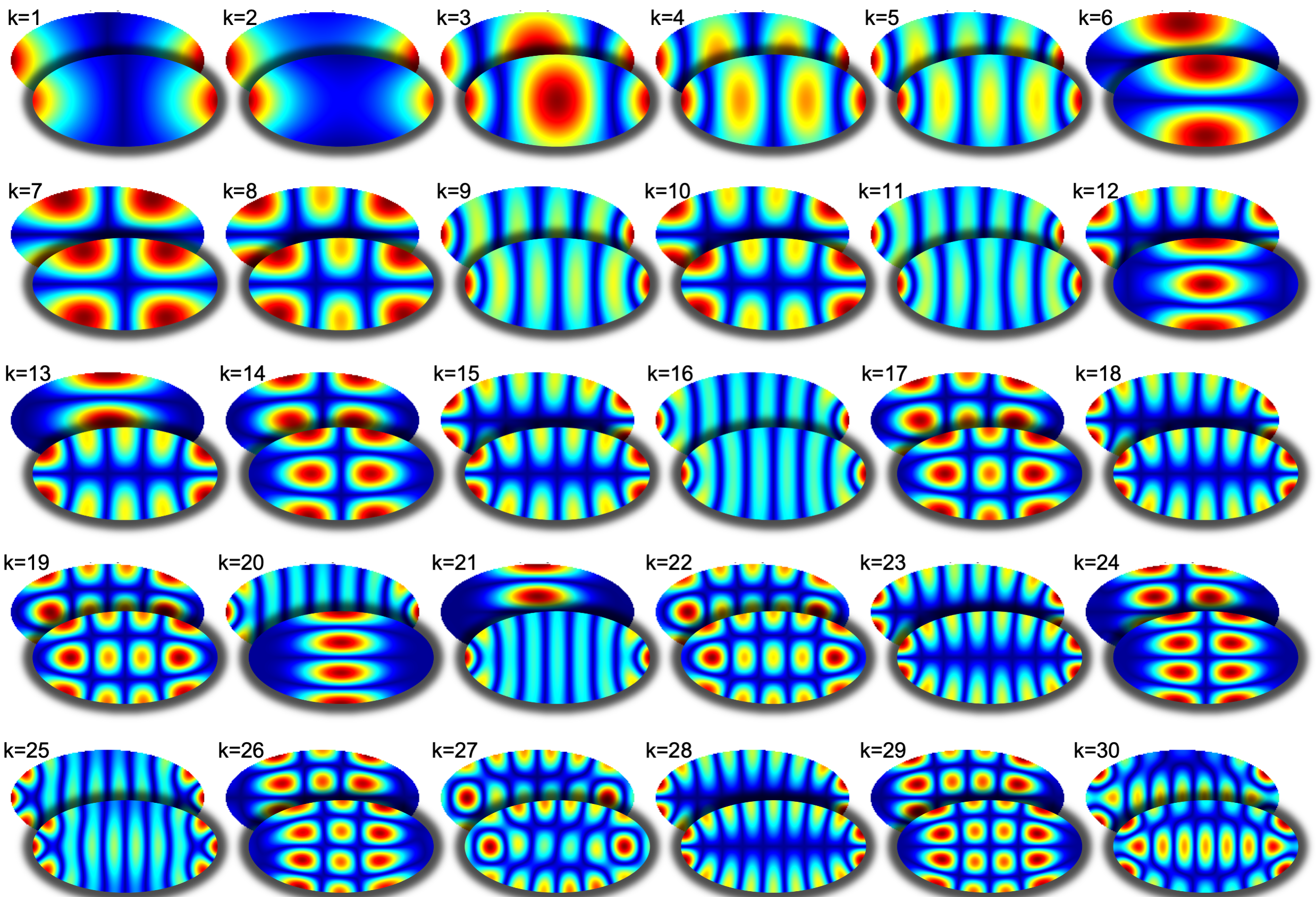}
\caption{\label{fig:ellipse_modes}Profiles of the first 30 magnetic eigenmodes of an ellipsoidal Permalloy nanoplatelet at zero external field, computed using FEM with 1nm mesh and FD method with cell size $1\times1\times 5$ nm$^3$. The mode patterns obtained with FEM are placed on top of those resulting from the FD simulations. Both methods yield identical mode patterns, with only minor differences in the frequencies (cf.~table \ref{tab:frequencies}) and occasional switches in the mode sequence (cf.~modes $12/13$ and $20/21)$.
The equilibrium configuration is a single-domain state with in-plane magnetization mainly oriented along the major axis.
 The color code displays the local oscillation strength, where red and blue denote maximum and minimum amplitude regions, respectively.}  
\end{figure*}

\section{Application Example}
We demonstrate the applicability of the methods detailed above on the example of a soft-magnetic thin-film nanoelement with an elliptical shape with a \SI{200}{\nano\meter} axis, \SI{100}{\nano\meter} minor axis, and \SI{5}{\nano\meter} thickness. The material parameters are those of Permalloy, i.e., ferromagnetic exchange constant $A=\SI{13}{\pico\joule\per\meter}$, spontaneous magnetization 
$M_s=\SI{8e5}{\ampere\per\meter}$, and zero magnetocrystalline anisotropy, $K_1=\SI{0}{\joule\per\meter\cubed}$. We assume a value of $\gamma=\SI{2.21e5}{\radian\meter\per\ampere\per\second}$ for the gyromagnetic ratio and consider the case where no static external field is applied.

In a first step, we calculate the relaxed zero-field magnetization structure---resulting in our case in a state with almost homogeneous magnetization which, due to shape anisotropy, is mainly oriented along the major axis direction. Using the methods described in section~\ref{sect:eigenmodes}, we numerically determine the eigenmodes and eigenfrequencies of this configuration and compare FEM and FD implementations. A selection of the results is shown in Fig.~\ref{fig:ellipse_modes}. 

\begin{figure}[ht]
  \includegraphics[width=\linewidth]{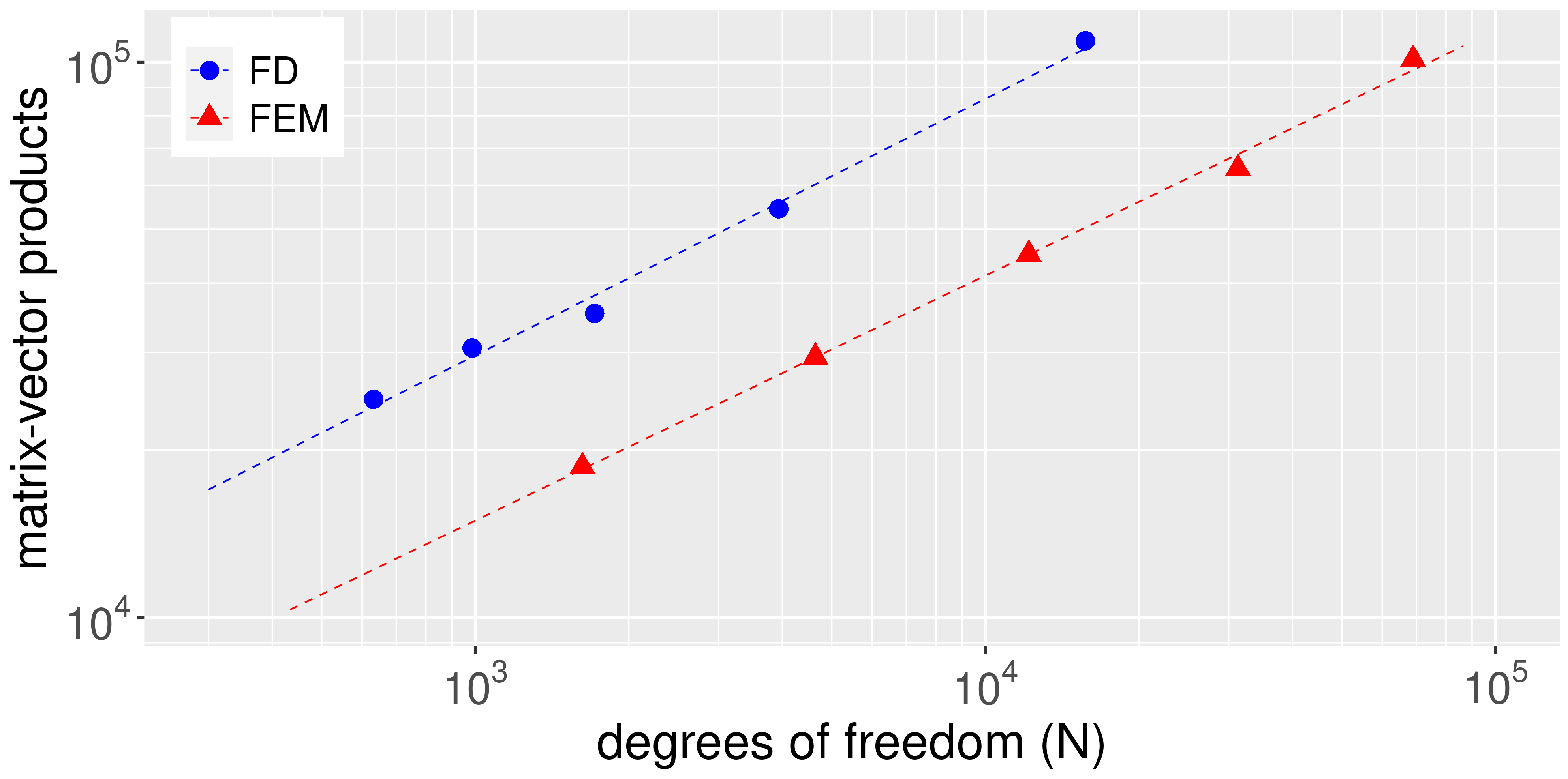}
    \caption{\label{cell_size_data}
    Number of matrix-vector product operations $A_{0\perp}\cdot\underline{v}$ required by the Arnoldi iterative method as a function of the number $N$ of degrees of freedom of the FD (blue) and FEM (red) implementations. }
\end{figure}

 We simulated this set of eigenmodes using a set of finite-element meshes of different discretization density, with cell sizes ranging between \SIrange{1}{5}{\nano\meter}, and obtained in all cases the same sequence of mode patterns, with minor differences in the resonance frequencies. The FEM results of Fig.~\ref{fig:ellipse_modes} were calculated using the mesh with \SI{1}{\nano\meter} cell size, consisting of more than 290\,000 irregularly shaped tetrahedral elements and about 69\,000 nodes, while the FD results were computed using 15708 prism cells with dimension $1\times 1\times 5$~nm$^3$.
  The computational cost of the eigensolver depends on the total number of matrix-vector product operations $A_{0\perp}\cdot\underline{v}$ required by the Arnoldi iterative method implemented in the ARPACK library. In Fig.~\ref{cell_size_data}, we show this quantity as a function of the number of degrees of freedom $N$ (i.e. number of cells for finite difference, number of nodes for finite element methods, respectively) for the computation of the first 30 eigenmodes. In these examples, the number of operations scales almost linearly and the computation time ranges between a few seconds for the smallest problem size in the FD formulation to several hours in the largest problem calculated with FEM.

We observe that there is substantial agreement between the results obtained with FD and FEM computations. The eigenfrequencies, 
reported in table \ref{tab:frequencies}, differ by a few percent in the worst case and display occasional inversion of eigenmodes sequence at very close frequency (this occurs, e.g., in FD modes 12/13, 20/21 compared to FEM modes 12/13, 20/21). We are not surprised about these small discrepancies since the FD method works well with sharp edges but 
cannot treat curved boundaries correctly. 
Both methods predict the fundamental (Kittel) mode close to the theoretical estimate for spatially-uniform magnetization based on demagnetizing factors $f_K=\gamma M_s/(2\pi)\sqrt{(N_z-N_x)(N_y-N_x)}\approx5.38$ GHz. 
We also note that the first two modes (\SI{5.07}{\giga\hertz} and \SI{5.08}{\giga\hertz} in FEM results, \SI{4.93}{\giga\hertz} and \SI{4.96}{\giga\hertz} in FD results) have almost identical profile regarding the oscillation amplitude. These two modes differ by their symmetry regarding the phase, which is not shown in this image. The lower-frequency mode refers to an anti-phase oscillation of the opposite ends, whereas these regions oscillate with the same phase in the mode with a slightly higher frequency.
\begin{table}[ht]
\begin{center}
\begin{tabular}{ccc}
\begin{tabular}{| c >{\columncolor{FEMGray}}c 
>{\columncolor{FDGray}}c |}
\hline
k & FEM & FD \\ \hline
1 & 5.07 & 4.93 \\
2 & 5.08 & 4.96 \\
3 & 6.35 & 6.32 \\
4 & 8.03 & 7.98  \\
5 & 10.15 & 10.09 \\
6 & 10.65 & 10.63 \\
7 & 11.46 & 11.37 \\
8 & 12.50 & 12.38 \\
9 & 12.56 & 12.50 \\
10 & 13.82 & 13.70 \\
\hline
\end{tabular}
\begin{tabular}
{| c  >{\columncolor{FEMGray}}c
>{\columncolor{FDGray}}c |}
\hline
k & FEM & FD \\ \hline
11 & 15.22 & 15.13 \\
12 & 15.34 & 15.34 \\
13 & 15.48 & 15.35 \\
14 & 16.65 & 16.63 \\ 
15 & 17.52 & 17.37 \\
16 & 18.13 & 17.99 \\
17 & 18.25 & 18.20 \\
18 & 19.91 & 19.78 \\
19 & 20.16 & 20.12 \\
20 & 21.08 & 21.08 \\
 \hline
\end{tabular}
\begin{tabular}
{| c  >{\columncolor{FEMGray}}c
>{\columncolor{FDGray}}c |}
\hline
k & FEM & FD \\ \hline

21 & 21.28 & 21.14 \\
22 & 22.40 & 22.37 \\
23 & 22.66 & 22.53 \\
24 & 22.91 & 22.97 \\
25 & 24.66 & 24.41 \\
26 & 25.04 & 25.02 \\
27 & 25.05 & 25.06 \\
28 & 25.74 & 25.57 \\
29 & 27.48 & 27.51 \\
30 & 27.94 & 27.81 \\
 \hline
\end{tabular}
\end{tabular}
\end{center}
\caption{Frequencies (in \si{\giga\hertz}) of the eigenmodes shown in Fig.~\ref{fig:ellipse_modes}, simulated with the finite-element method (FEM) and the finite-difference method (FD).
\label{tab:frequencies}}
\end{table}
  
The eigenmodes and their frequencies characterize the dynamic system insofar as they represent the fundamental patterns at which the magnetization can oscillate around the equilibrium state. If driven by a harmonic external field, the system oscillates at the applied field's frequency, which generally is not equal to one of the system's eigenfrequencies. Nevertheless, the resulting steady-state pattern of the dynamic magnetization can be decomposed into a superposition of eigenmodes. The strength at which an eigenmode is excited depends on how close its eigenfrequency is to that of the applied field. However, one should note that only a subset of the possible eigenmodes is usually excited, as not all magnetic eigenoscillations are compatible with the applied field's spatial distribution and orientation. 

\begin{figure}[ht]
\includegraphics[width=\linewidth]{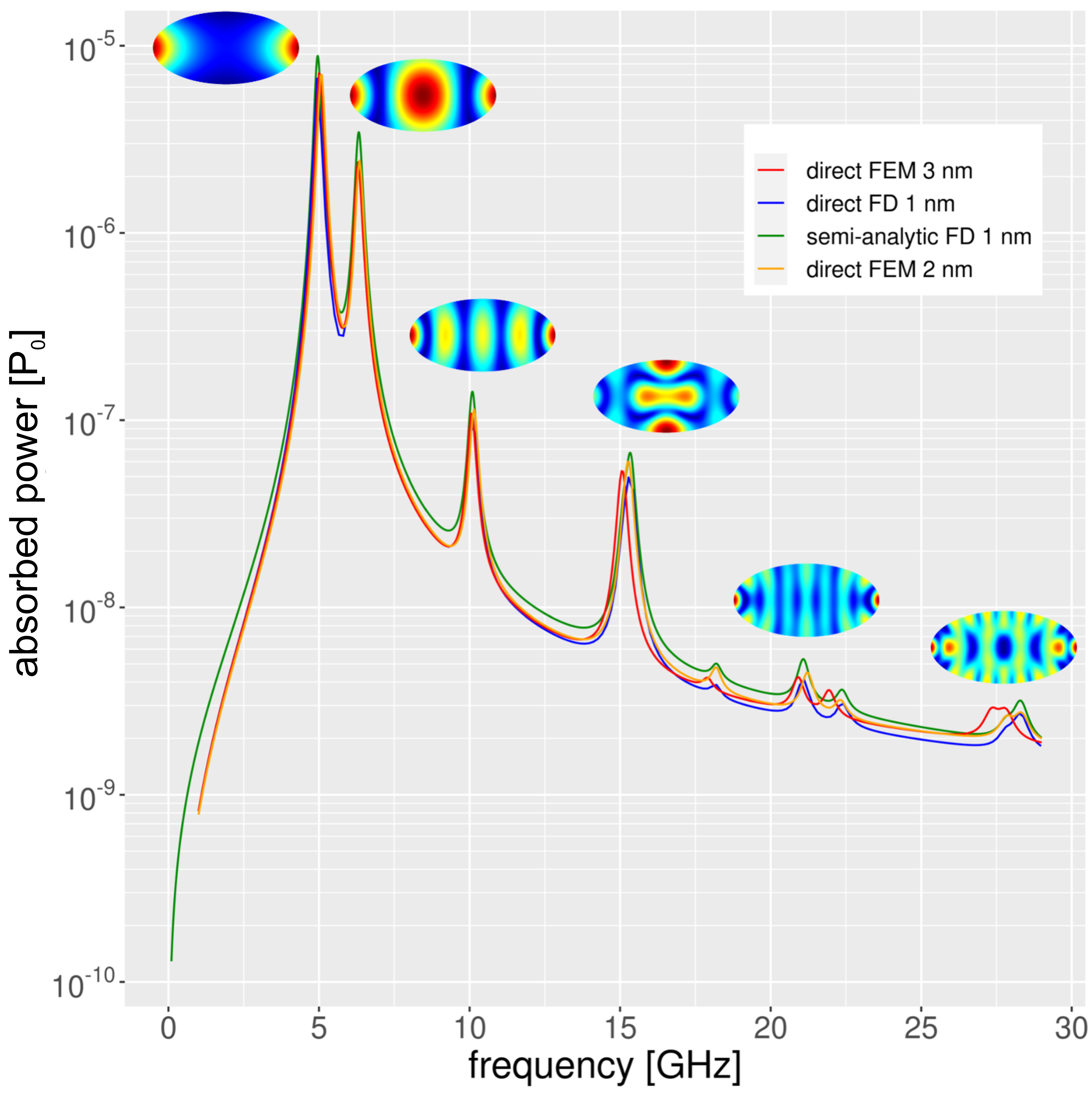}
\caption{\label{fig:ellipse_scan} Magnetic absorption spectrum and mode patterns at the peaks generated by an applied rf field. The response of the magnetic system depends strongly on the external field's frequency. The oscillation patterns at the dominant absorption peaks can be clearly identified as the lowest-frequency eigenmodes shown in Fig.~\ref{fig:ellipse_modes}, with the peaks' frequencies corresponding to those of specific eigenmodes. The lines labeled as ``direct'' refer to solutions of eq.~\eqref{eq:discrete forced LLG response} via MF-$\mu$LRS plugged into eq.~\eqref{eq:discrete absorbed power} and taking the real part, while ``semi-analytic'' refers to eqs.~\eqref{eq:compact discrete absorbed power}-\eqref{eq:compact discrete absorbed active power}. 
The absorbed power $P_\text{abs}$ is measured in  units of $\mu_0  M_s^2 \cdot V  \cdot \gamma  M_s$, which in our example corresponds to  about \SI{11.2}{\micro\watt}.
} 
\end{figure}

To illustrate the rf-field-driven dynamics and its connection to eigenmodes, we investigate the frequency-dependent linear response to an externally applied sinusoidal magnetic field of low amplitude. Specifically, we apply a time-harmonic field with amplitude \SI{0.5}{\milli\tesla} along the $y$-axis (the minor axis direction), assume a Gilbert damping constant $\alpha=0.01$, and simulate the frequency-dependent stationary magnetization dynamics developing in the applied rf field. The frequency of the field is increased in steps of \SI{50}{\mega\hertz} in a range from 
\SIrange{0.05}{30}{\giga\hertz}. At each frequency, we solve the system \eqref{eq:discrete forced LLG response} and thereby obtain the frequency-dependent profile of the dynamic magnetization.

Fig.~\ref{fig:ellipse_scan} shows the simulated power absorption driven by the magnetic rf field, computed according to eq.~\eqref{eq:discrete absorbed power} using both FD and FEM solvers. 
As in the case of the eigenmode calculations, very similar results are obtained with both FEM and FD, which moreover vary only insignificantly with changes of the cell size between about one and three \si{\nano\meter}. We also report the diagram resulting from the application of semi-analytical formulas  \eqref{eq:compact discrete absorbed power}-\eqref{eq:compact discrete absorbed active power} derived in section \ref{sect:semi-analytic} based on previously computed eigenmodes. These results are similar to those that can be computed by evaluating the imaginary part of the spatially-averaged dynamical susceptibility~\cite{labbe_microwave_1999,vukadinovic_magnetic_2000, dmytriiev_calculation_2012}.

Distinct peaks can be identified in the frequency dependence of the absorbed power, with the most pronounced ones being near \SI{5.0}{\giga\hertz} and \SI{6.3}{\giga\hertz}. The frequency of the first absorption peak is very close to that of the first and second eigenmodes, as listed in table~\ref{tab:frequencies}. An inspection of the dynamic magnetization unfolding near \SI{5.0}{\giga\hertz} shows that only the symmetric oscillation of mode \#2 is excited in this setup, not the antisymmetric one of mode \#1. The second resonance  in the spectrum,
at \SI{6.3}{\giga\hertz}, can be identified as the third eigenmode $k=3$ shown in Fig.~\ref{fig:ellipse_modes} in terms of both the mode pattern and frequency, as can be seen from the mode profiles in the insets of Fig.~\ref{fig:ellipse_scan}.
Similarly, the oscillation pattern at \SI{10.0}{\giga\hertz} can be ascribed to the fifth eigenmode $k=5$ shown in Fig.~\ref{fig:ellipse_modes}, which has a nominally identical eigenfrequency according to table~\ref{tab:frequencies}. However, the fourth mode, expected near \SI{8.0}{\giga\hertz}, does not appear as a peak in the spectrum, indicating that it cannot be excited by an oscillating field in the $y$ direction. 
Furthermore, the intensity of the resonances diminishes significantly as the frequency increases (note the logarithmic scale).
Correlating the dynamic magnetization patterns of the higher-frequency peaks at about
\SI{15.0}{\giga\hertz} and \SI{21.0}{\giga\hertz}  to specific mode profiles  shown in Fig.~\ref{fig:ellipse_modes} is less evident than for the first three principal peaks. The mode pattern of the resonance at \SI{15.0}{\giga\hertz} appears to be a superposition of at least two nearby eigenmodes, $k=11$ and $k=12$, that have similar eigenfrequencies. The mode pattern in the vicinity of \SI{21.0}{\giga\hertz} resembles that of eigenmode \#21. Finally, although the complex pattern developing at the weak absorption peak around \SI{27.5}{\giga\hertz} does not correspond to any of the previously determined eigenmodes, one can suspect that its main features are reproduced by a superposition of the modes $k=29$ and $k=30$.

These results, particularly as far as the dominant resonances are concerned, exemplify the principle discussed before that the driven stationary dynamics, at any frequency, can be understood as a superposition of eigenmodes. Note that, when simulating the data labeled as ``direct'' in Fig.~\ref{fig:ellipse_scan}, we solved eq.~(\ref{eq:discrete forced LLG response}) without using the knowledge of the eigenmodes, which served only to identify and interpret the resulting resonances. The formalism discussed in section \ref{sect:semi-analytic} provides a rigorous framework for such a decomposition into eigenmodes.
By representing the dynamic magnetization as a linear combination of eigenmodes and furnishing the frequency-dependent coefficients of such an expansion through eq.~(\ref{eq:approx discrete forced LLG response series}), the semi-analytic method allows us to calculate results that are almost identical to those of the direct numerical solution (as illustrated by the green line in fig.~\ref{fig:ellipse_scan}), but it can do so in a much faster way---in fact almost instantaneously---once the eigenmodes and eigenfrequencies are computed.
Needless to say that magnetization frequency response \eqref{eq:approx discrete forced LLG response series}-\eqref{eq:approx reduced discrete forced LLG response} and power absorption spectra \eqref{eq:approx power spectrum}-\eqref{eq:approx discrete absorbed active power} can be evaluated for any (small) damping and rf-field without making further simulations. 

\section{Conclusion}
Scientific progress in magnonics confers growing importance on simulation studies in this domain and calls for efficient and precise numerical methods to simulate systems that can be directly compared to experiments. In this article, we presented different numerical approaches to determine the high-frequency dynamics of micromagnetic systems while observing the major computational imperatives of short calculation times and low memory requirements. Linearizing the LLG equation and solving for the relevant equations in the frequency domain allows obtaining precise results on the high-frequency dynamics of micromagnetic systems in a significantly faster way compared to the traditional and much more tedious approach, which consists in calculating the magnetization dynamics in the time domain and then Fourier-analyzing it. The remarkable gain in speed of these frequency-domain simulations opens the way towards extensive numerical studies of magnonic systems on, e.g., the systematic impact of external parameters over a broad range, such as gradually varying external fields\cite{cheenikundil2022high}. 

We demonstrated that the presented methods are independent of the discretization type, yielding the same results in FEM and FD formulations. Owing to a fully sparse, operator-based implementation, they allow for large-scale simulations of the dynamics in the frequency domain with essentially identical memory requirements as modern micromagnetic algorithms operating in the time domain. 
The semi-analytic approach discussed in section \ref{sect:semi-analytic} and confirmed by numerical results is particularly powerful as it gives the possibility to immediately determine the complete frequency-dependent response of a system solely based on the system's eigenmodes and eigenvalues without solving a system of equations at each frequency. 

These numerical methods have such significant benefits in speed and accuracy compared to classical time-domain micromagnetic simulations that we anticipate they will become a {\em de facto} standard for the modeling of magnonic systems.

\begin{acknowledgments}
RH acknowledges the High Performance Computing center of
the University of Strasbourg for supporting this work by providing access to computing resources.

\end{acknowledgments}

%\bibliography{references}
%\end{document}

\end{document}